\documentclass[sigconf]{acmart}
\usepackage{hyperref}
\usepackage{xcolor}
\usepackage{multirow} 
\usepackage{stfloats} 

\AtBeginDocument{%
  }

\setcopyright{acmlicensed}
\copyrightyear{2018}
\acmYear{2018}
\acmDOI{XXXXXXX.XXXXXXX}
\acmConference[Conference acronym 'XX]{Make sure to enter the correct
  conference title from your rights confirmation email}{June 03--05,
  2018}{Woodstock, NY}
\acmISBN{978-1-4503-XXXX-X/2018/06}

\begin{document}

\title[Data-Centric Multilingual E-Commerce Search]{A Data-Centric Approach to Multilingual E-Commerce Product Search: Case Study on Query-Category and Query-Item Relevance}

\author{Yabo Yin}
\email{yinyabo22@outlook.com}
\affiliation{%
  \institution{No Affiliation}
  \country{China}
  }
  
\author{Yang Xi}
\email{23140666@bjtu.edu.cn}
\affiliation{%
  \institution{Beijing Jiaotong University}
  \country{China}
  }
\author{Jialong Wang}
\email{wjl_906@xauat.edu.cn}
\affiliation{%
  \institution{Xi’an University of Architecture and Technology}
  \country{China}
  }
\author{Shanqi Wang}
\email{19917805190@163.com}
\affiliation{%
  \institution{Harbin University of Science and Technology}
  \country{China}
  }
\author{Jiateng Hu}
\email{1178183680@qq.com}
\affiliation{%
  \institution{Cangzhou Jiaotong College}
  \country{China}
  }

\renewcommand{\shortauthors}{Yin et al.}
\begin{abstract}
Multilingual e-commerce search suffers from severe data imbalance across languages, label noise, and limited supervision for low-resource languages—challenges that impede the cross-lingual generalization of relevance models despite the strong capabilities of large language models (LLMs). In this work, we present a practical, architecture-agnostic, data-centric framework to enhance performance on two core tasks: Query-Category (QC) relevance (matching queries to product categories) and Query-Item (QI) relevance (matching queries to product titles). Rather than altering the model, we redesign the training data through three complementary strategies: (1) translation-based augmentation to synthesize examples for languages absent in training, (2) semantic negative sampling to generate hard negatives and mitigate class imbalance, and (3) self-validation filtering to detect and remove likely mislabeled instances. Evaluated on the CIKM AnalytiCup 2025 dataset, our approach consistently yields substantial F1 score improvements over strong LLM baselines, achieving competitive results in the official competition. Our findings demonstrate that systematic data engineering can be as impactful as—and often more deployable than—complex model modifications, offering actionable guidance for building robust multilingual search systems in the real-world e-commerce settings.
\end{abstract}

\ccsdesc[500]{Information systems~Retrieval tasks and goals}
\ccsdesc[300]{Information systems~Multilingual and cross-lingual retrieval}
\ccsdesc[300]{Computing methodologies~Natural language processing}
\ccsdesc[100]{Applied computing~E-commerce}

\keywords{Multilingual E-commerce Search,  Query-Category Relevance, Query-Item Relevance, Large Language Models, Data Augmentation}

\maketitle
\section{Introduction}
The rapid expansion of global e-commerce has led to an increasingly multilingual and multicultural user base on platforms such as AliExpress, Lazada, and Daraz. In such environments, effective product search is not merely a convenience—it is a core driver of user satisfaction, conversion rates, and platform competitiveness. Central to this process are two fundamental tasks: Query-Category (QC) relevance \cite{zhu2023hcl4qc}, which maps a user’s search query to the most appropriate product category path, and Query-Item (QI) relevance \cite{meng2025query}, which determines whether a specific product matches the user’s intent. Both tasks require a deep understanding of user intent, multilingual semantics, and domain-specific knowledge.

The advent of large language models (LLMs) has significantly advanced capabilities in natural language understanding, enabling strong performance across a wide range of tasks \cite{achiam2023gpt,workshop2022bloom,touvron2023llama}. When adapted via supervised instruction tuning \cite{bai2022training}, LLMs have become a de facto standard baseline for many relevance modeling tasks due to their simplicity and competitive performance. However, their effectiveness in real-world multilingual e-commerce settings remains constrained by data-related challenges.

A key limitation, as evidenced by the CIKM AnalytiCup 2025 Multilingual E-Commerce Search Challenge dataset, is the severe data imbalance across languages. While English, French, and Spanish are well represented in the training set, other languages—such as Italian, Polish, and Arabic—are entirely absent from the QC and QI tasks.
This gap impedes cross-lingual generalization and undermines the model’s ability to serve a truly global user base equitably. Compounding this issue are label imbalance and annotation noise, which introduce biases that particularly hurt performance on hard negatives and fine-grained relevance distinctions.

Prior work \cite{zhang2021modeling, zhu2023hcl4qc, wu2024cprm} on QC and QI relevance has often pursued performance gains through model-centric innovations—such as architectural modifications, domain-adaptive pretraining, or sophisticated fine-tuning strategies. While effective, these approaches typically require task-specific engineering and may not generalize easily across settings. In contrast, we argue that a systematic, data-centric perspective offers a more practical and model-agnostic pathway to robust multilingual relevance modeling. In this paper, we present a purely data-centric enhancement framework developed for the CIKM AnalytiCup 2025 competition. Rather than modifying the model architecture, we focus on redesigning the training data pipeline to directly address the dual challenges of language coverage and label quality. Our approach integrates three complementary data engineering strategies: 
\textbf{(1) Translation-Based Augmentation}: We leverage a high-quality massively multilingual translation model to synthesize training examples for underrepresented or missing languages, enabling cross-lingual knowledge transfer without requiring native supervision;
\textbf{(2) Semantic Negative Sampling}: To improve relevance discrimination, we construct hard negative examples by pairing translated queries with semantically plausible but incorrect category paths or items, thereby sharpening the model’s decision boundaries;
\textbf{(3) Self-Validation Filtering}: We employ model-based consistency checks across multiple inference passes to identify and remove likely mislabeled or ambiguous training instances, enhancing overall label reliability.


We apply this framework to both QC and QI tasks using Qwen3 \cite{yang2025qwen3} as the base LLM, kept fixed throughout. Extensive experiments demonstrate that our data-centric enhancements lead to consistent F1 score improvements on both tasks, outperforming baseline models trained on original data. Our main contributions are as follows:
\begin{itemize} 
    \item We propose a practical, model-agnostic, data-centric framework for multilingual e-commerce relevance tasks, explicitly targeting language imbalance and label noise.
    \item We demonstrate how translation-based augmentation and semantic negative sampling can enhance cross-lingual generalization and fine-grained relevance modeling without additional human annotation.
    \item We provide actionable insights for practitioners, showing that data-focused interventions can be as impactful as, and often more deployable than, complex model modifications. 

\end{itemize}

\section{Related Work}

E-commerce search constitutes a specialized domain within information retrieval, focusing on the matching of user queries with relevant products or categories—a task commonly referred to as semantic matching or relevance ranking \cite{barrera2016enhancing}. The multilingual context of global e-commerce platforms further complicates this task, requiring systems to process diverse languages while preserving consistent relevance standards across different regions.

Early multilingual e-commerce search systems often relied on cross-lingual word embeddings \cite{ahuja2020language} or machine translation techniques to bridge linguistic gaps \cite{zhang2023machine}, typically by projecting all languages into a unified semantic space \cite{maimaiti2025improving}. More recently, the emergence of large language models (LLMs) has spurred new lines of research. For instance, \citet{kumar2024enhancing} demonstrated how transformer-based LLMs can automate the generation of multilingual product descriptions, thereby enhancing content localization and customer engagement. In addressing cold-start scenarios prevalent in emerging markets, \citet{wang2025csrm} introduced CSRM-LLM, a framework that leverages cross-lingual transfer learning via machine translation and retrieval-augmented query expansion. For cross-lingual information retrieval, \citet{mcnabb2025llm} developed an LLM-enhanced machine translation system tailored to domain-specific challenges such as polysemy and non-translatable entities in e-commerce queries.

LLMs have also shown significant potential in relevance prediction tasks. \citet{tang2025lref} proposed LREF, an LLM-based relevance framework that integrates supervised fine-tuning with data selection, chain-of-thought reasoning, and direct preference optimization to mitigate issues like over-recall and reasoning inconsistency. Similarly, \citet{duraj2025chain} employed chain-of-thought reasoning in LLMs to improve query-category alignment, underscoring the value of reasoning-enhanced methods in complex e-commerce taxonomies.

Despite these advances, the majority of existing studies adopt a model-centric approach, prioritizing architectural innovations or advanced fine-tuning strategies. In contrast, our work adopts a data-centric perspective, focusing on systematic data engineering to address persistent challenges such as language imbalance and label quality in multilingual e-commerce search.

\section{Method}

\begin{figure}[!t]
\centering
\includegraphics[width=0.99\linewidth]{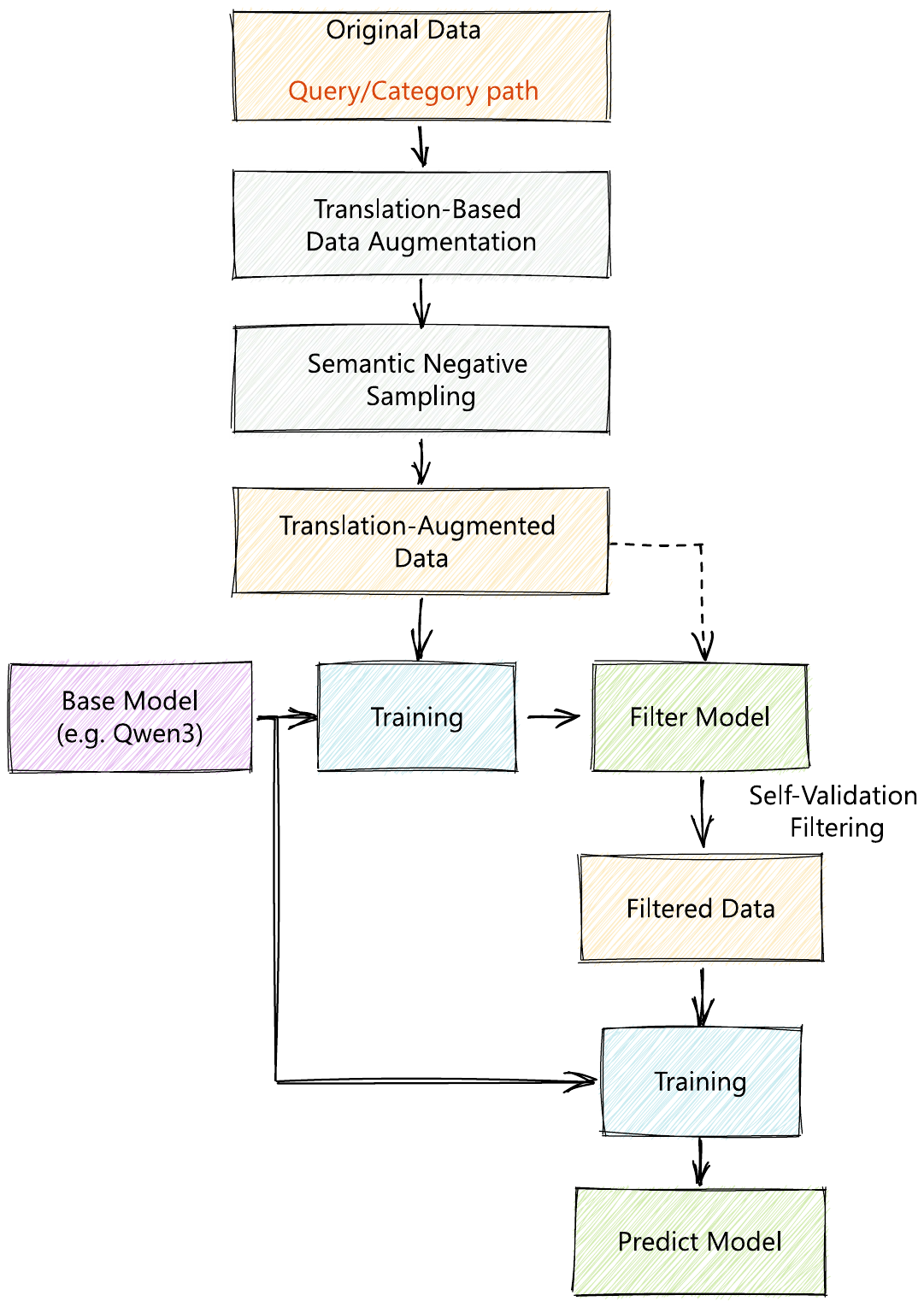}
\caption{
Overview of the proposed data-centric framework for the Query-Category (QC) relevance task.
}
\label{fig:overall}
\end{figure}

We propose a data-centric framework to enhance the performance of LLMs on multilingual e-commerce relevance tasks. The core of our approach lies in systematically engineering the training data to address the specific challenges of language generalization, class imbalance, and label noise, rather than modifying the underlying model architecture. An overview of the framework for the Query-Category (QC) task is illustrated in Figure \ref{fig:overall}, and a similar pipeline is applied to the Query-Item (QI) task.

\subsection{Base Model and Task Formulation}
We employ Qwen3-14B-Base \cite{yang2025qwen3} as our foundation model for both tasks. This model was selected for its strong multilingual capabilities, demonstrated performance across diverse NLP benchmarks, and extensive pre-training corpus that covers all languages present in the competition.
We formulate both QC and QI tasks as binary classification problems.
The model is fine-tuned to generate either "yes" or "no" in response to a natural language prompt that encapsulates the query and the candidate (either a category path or an item title), similar to Qwen3-Reranker \cite{zhang2025qwen3_embedd}. We adopt an Alpaca-style instruction-tuning format, which has been demonstrated to effectively align model outputs with task requirements. 

\subsection{Data-Centric Enhancement Strategies }

\subsubsection{\textbf{Translation-Based Data Augmentation for Language Generalization}}
The original training data for the QC/QI task lacks examples for four languages present in the development and test sets: German, Italian, Polish, and Arabic. To bridge this zero-shot generalization gap, we synthesize training data for these languages.
We use the ByteDance-Seed/Seed-X-PPO-7B \cite{cheng2025seedxbuildingstrongmultilingual} model, a massively multilingual translation model supporting 28 languages, to translate the original query from existing high-resource language training samples (e.g., Spanish) into the target low-resource languages (e.g., German), while keeping the corresponding category paths (for QC) or item titles (for QI) and their labels unchanged. This process generates synthetic samples for each missing language, effectively balancing the language distribution in the training set and enabling cross-lingual knowledge transfer.

\subsubsection{\textbf{Semantic Negative Sampling for Class Imbalance}}
Analysis of the original training data for both QC and QI tasks reveals a significant class imbalance, with positive samples (label=1) substantially outnumbering negative ones (label=0). This imbalance can lead to models with poor discrimination capabilities, as they may learn to prioritize the majority class without developing robust feature representations for distinguishing between relevant and irrelevant pairs. To address this limitation, we augment the training dataset with synthetically generated hard negative samples.

For each target language, we begin by sampling positive examples from other languages. The procedure for generating hard negatives is as follows:
\begin{itemize}
\item \textbf{Query Translation:}  The query from the sampled positive example is translated into the target language using the same multilingual translation model described previously.
\item \textbf{Hard Negative Candidate Sampling:} For the QC task, we employ the Qwen3-Embedding-4B model to encode all category paths into dense vector representations. For a given positive sample's original category path, we retrieve the top-20 to top-50 most semantically similar category paths from the entire category set based on cosine similarity in the embedding space. From this retrieved set, we randomly sample one category path to replace the original positive category, thereby constructing challenging negative examples where the category is semantically related to the query but ultimately incorrect. For the QI task, we apply a similar strategy for item titles.
\item \textbf{Negative Label Assignment:} The newly constructed pair (translated query, sampled negative category/item) is assigned a negative label of 0.
\end{itemize}
This semantic negative sampling strategy creates challenging negative examples that are semantically proximate to positive instances, forcing the model to learn finer-grained distinctions and develop more robust decision boundaries for relevance classification.

\subsubsection{\textbf{Self-Validation Filtering for Noisy Labels}}

In preliminary experiments on both QC and QI tasks, we identified noticeable label noise in the training data, a situation further exacerbated by our data synthesis strategies. Such label inaccuracies—stemming from either human annotation errors or biases introduced during synthetic data generation—are prevalent in real-world applications. To mitigate the impact of label noise, we implement a self-validation filtering strategy.

First, we fine-tune Qwen3-14B-Base on the processed training data to obtain an initial filter model. This model is then used to infer over the full training set, flagging samples where high-confidence predictions contradict original labels as potential annotation errors.  
We subsequently apply additional processing to these identified samples. Potential actions include removing high-confidence mislabeled instances, performing human verification, or employing multiple robust models (such as DeepSeek-R1 \cite{deepseekai2025deepseekr1incentivizingreasoningcapability} and Qwen3-Max \cite{yang2025qwen3}) as judges to conduct majority-vote label correction. In this work, to maintain cost efficiency, we opt to simply remove high-confidence mislabeled samples. 
Finally, we train the final prediction model on the filtered or corrected training dataset and use this model for all subsequent evaluations and test set predictions.

\subsection{Model Training and Inference}
\subsubsection{\textbf{Model Training}}
For both tasks, we fine-tune the Qwen3-14B-Base model using Low-Rank Adaptation (LoRA) \cite{hu2022lora} to enhance training efficiency. The model is optimized using the standard Supervised Fine-Tuning (SFT) objective, defined as:
\begin{equation}
        \mathcal{L} = -\log p(l\;|\;P(q,c)),
\end{equation}
where $P(\cdot,*)$ denotes the probability assigned by the LLM, and $l$ is the target label—either “yes” for positive pairs or “no” for negative pairs. Here, $q$ represents the user query, while $c$ denotes the candidate category path in the QC task or the item description in the QI task. This loss function encourages the model to maximize the likelihood of the correct label, thereby improving predictive accuracy.

\subsubsection{\textbf{Model Inference}}
During inference, we utilize the vLLM \cite{kwon2023efficient} framework to enable high-throughput generation. Rather than directly using the raw “yes”/“no” output, we compute the probability of the model generating “yes” by normalizing the log-probabilities of the two output tokens:
\begin{equation}
    p(\text{yes}) = \frac{e^{p(\text{yes})}}{e^{p(\text{yes})} + e^{p(\text{no})}}
 \end{equation}

A task-specific threshold is then applied to this probability to obtain the final binary prediction:
\begin{equation}
    \text { pred }=\left\{\begin{array}{ll}
\text { yes, } & \text { if } p(\text { yes }) \geq \text { threshold } \\
\text { no, } & \text { if } p(\text { yes })<\text { threshold }
\end{array}\right.
\end{equation}

\section{Experiments}
\subsection{Experimental Settings}
\subsubsection{{Datasets} }

\begin{table*}[!t]
\centering
\caption{Dataset Statistics for QC and QI Tasks}
\label{tab:transposed_data}
\begingroup
\resizebox{0.9\linewidth}{!}{  
\begin{tabular}{c|c|ccccccccccccc} 
\toprule 
\multicolumn{2}{c|}{\textbf{Task}} & English & French & Spanish & Korean & Portuguese & Japanese & German & Italian & Polish & Arabic & Thai & Vietnamese & Indonesian \\
\midrule
\multirow{3}{*}[-0.2ex]{\textbf{QC}} & \textbf{Train}    & 50k     & 50k    & 50k     & 50k    & 50k        & 50k      & --     & --      & --     & --     & --   & --         & --         \\
~ & \textbf{Dev}      & 10k     & 10k    & 10k     & 10k    & 10k        & 10k      & 10k    & 10k     & 10k    & 10k    & --   & --         & --         \\
~ & \textbf{Test}     & 10k     & 10k    & 10k     & 10k    & 10k        & 10k      & 10k    & 10k     & 10k    & 10k    & --   & --         & --         \\
\midrule 
\multirow{3}{*}[-0.2ex]{\textbf{QI}} & \textbf{Train}    & 40k     & 40k    & 40k     & 45k    & 40k        & 45k      & -    & -     & -    & -    & 40k  & -        & -       \\
~ & \textbf{Dev}      & 10k     & 5k     & 5k      & 5k     & 5k         & 5k       & 5k     & 5k      & 5k     & 5k     & 5k   & 5k         & 5k         \\
~ & \textbf{Test}     & 10k     & 10k    & 10k     & 10k    & 10k        & 10k      & 10k    & 10k     & 10k    & 10k    & 10k  & 10k        & 10k        \\
\bottomrule
\end{tabular}
}
\endgroup
\end{table*}

We evaluate our method on the dataset from the CIKM AnalytiCup 2025 Multilingual E-Commerce Search Competition, which comprises two core tasks:
\begin{itemize} 
    \item  Query-Category Relevance (QC): Determine if a user's query is relevant to a given product category path (e.g., "Electronics > Audio Devices > Headphones").
    \item Query-Item Relevance (QI): Determine if a candidate item's title is relevant to a user's query.
\end{itemize}
The dataset exhibits real-world challenges: multilingual queries across 10+ languages, uneven language distribution in the training set (with some languages entirely absent in QC training), and inherent class imbalance. The statistics of the original datasets are detailed in Table \ref{tab:transposed_data}.

\subsubsection{Implementation Details}

We employ \textbf{Qwen/Qwen3-14B-Base} as the backbone model for both tasks, leveraging its robust multilingual pretraining foundation. All models are fine-tuned using Low-Rank Adaptation (LoRA) \cite{hu2022lora} with a rank of 32 and a dropout rate of 0.1, implemented through the LLaMA-Factory framework\footnote{https://github.com/hiyouga/LLaMA-Factory} \cite{zheng2024llamafactory}.

For the QC relevance task, input sequences are formatted using an Alpaca-style \cite{taori2023alpaca} instruction template that effectively contextualizes the query, category path, and user language information. The model is trained with a batch size of 128, learning rate of $7.5\times10^{-5}$, and cosine learning rate scheduler over a single epoch.  During inference, we utilize vLLM\footnote{https://github.com/vllm-project/vllm} \cite{kwon2023efficient} for efficient generation, with final predictions derived from normalized log-probabilities of "yes"/"no" responses using a calibrated decision threshold of 0.4.

For the QI relevance task, the instruction template incorporates item titles and is designed to handle spelling variations and intent ambiguity. After filtering mislabeled instances, we train the model over 2 epochs with a batch size of 64, learning rate of $5\times10^{-5}$, and maximum sequence length of 512 tokens. The prediction threshold is set to 0.2.

All experiments are conducted on 4 * RTX 4090 GPUs with bfloat16 precision and FlashAttention-2 enabled. 

\subsubsection{{ Evaluation Metrics} }
Both tasks are treated as binary classification problems (1 = relevant, 0 = irrelevant). Performance is measured by the F1 score on positive (relevant) samples. The final  metric is the average of the two task F1 scores: 
 
\begin{equation}
    F1^{\text {avg }}=\frac{F1^{\text {QC  }} + F1^{\text {QI }}}{2}
\end{equation}
where F1 for each task is calculated by:
\begin{equation}
   F1=2 \times \frac{\text { precision } \times \text { recall }}{\text { precision }+ \text { recall }}
\end{equation}

\subsection{{Experimental Results}}
\begin{table*}[!t]
\begin{center}
    \caption{Performance comparison of different data-centric strategies on the dev set (preliminary competition phase). Results are reported in F1 score for Query-Category (QC) and Query-Item (QI) relevance tasks. Best results are in bold.}
    \begingroup
    \resizebox{0.9\linewidth}{!}{%
    \begin{tabular}{ c| c | l | c | c }
    \toprule
    \multirow{2}{*}[-0.75ex]{Base Model} &
    \multirow{2}{*}[-0.75ex]{Training Data Type} &
    \multicolumn{1}{c|}{\multirow{2}{*}[-0.75ex]{Method}} &  
    \multicolumn{2}{c}{TASK} \\
    \cmidrule(lr){4-5}
    ~ & ~ & ~ & QC & QI \\
    \midrule
    Qwen3-8B  & Original   &  SFT                                                                      & 0.8733 & 0.8537  \\
    Qwen3-8B  & Original   &  DPO                                                                      & 0.8765 & -      \\
    \midrule
    
    Qwen3-14B & Original   &  SFT                                                                      & 0.8760 & 0.8662  \\
    Qwen3-14B & Augmented  &  SFT  + \textbf{Translation Augmentation}                                                    & 0.8834 & 0.8738  \\
    Qwen3-14B & Augmented  &  SFT  + \textbf{Translation Augmentation + Semantic Negative Sampling  }                        & 0.8885 & -       \\
    Qwen3-14B & Augmented  &  SFT  + \textbf{Translation Augmentation + Semantic Negative Sampling + Self-Validation Filtering }               & 0.8873 & -       \\
    
    Qwen3-14B & Augmented  &  SFT  + \textbf{Translation Augmentation + Semantic Negative Sampling + Task-Specific Threshold } & \textbf{0.8919} & -       \\
    Qwen3-14B & Augmented  &  SFT  + \textbf{Translation Augmentation + Self-Validation Filtering }                          & -      &  0.8799 \\
    Qwen3-14B & Augmented  &  SFT  + \textbf{Translation Augmentation + Self-Validation Filtering + Task-Specific Threshold }                 & -      &  \textbf{0.8839} \\
    \bottomrule
    \end{tabular}
    }
    \endgroup
    \label{tab:task_qc_qi_extended}
\end{center}
\end{table*}

We conduct comprehensive experiments to evaluate our proposed data-centric enhancement strategies. While certain configurations (marked with "-" in Table \ref{tab:task_qc_qi_extended}) remain unexplored due to competition time constraints, the obtained results offer valuable insights for data-centric approaches in multilingual e-commerce applications.

Our baselines established with standard supervised fine-tuning show expected scaling effects: Qwen3-8B achieves F1 scores of 0.8733 (QC) and 0.8537 (QI), while Qwen3-14B improves to 0.8760 (QC) and 0.8662 (QI). The consistent performance gains from translation-based data augmentation—improving QC from 0.8760 to 0.8834 and QI from 0.8662 to 0.8738—highlight the importance of mitigating zero-shot generalization gaps in multilingual settings.

Different tasks benefit from distinct enhancement strategies. For QC, incorporating semantic negative sampling with translation augmentation yields further improvement (0.8885 F1), suggesting that class imbalance mitigation through hard negatives is particularly valuable for category relevance tasks. For QI, combining translation augmentation with self-validation filtering achieves 0.8799 F1, indicating that label noise correction plays a more critical role in query-item relevance assessment, potentially due to its more subjective nature.

The optimal configurations—combining multiple strategies with task-specific threshold optimization—achieve 0.8919 F1 for QC and 0.8839 F1 for QI. These findings suggest that successful data-centric approaches should be tailored to task characteristics, with different data challenges requiring prioritized attention across different e-commerce relevance tasks.

Our work demonstrates that systematic data engineering, even without model architecture modifications, can effectively address common challenges in multilingual e-commerce applications. The task-dependent effectiveness of various strategies provides practical insights for future data-centric approaches in similar domains, emphasizing the need for diagnostic analysis of data characteristics before implementing enhancement strategies.

\subsection{Effectiveness Analysis of Task-Specific Threshold Setting}

\begin{figure}[!h]
\centering
\includegraphics[width=0.99\linewidth]{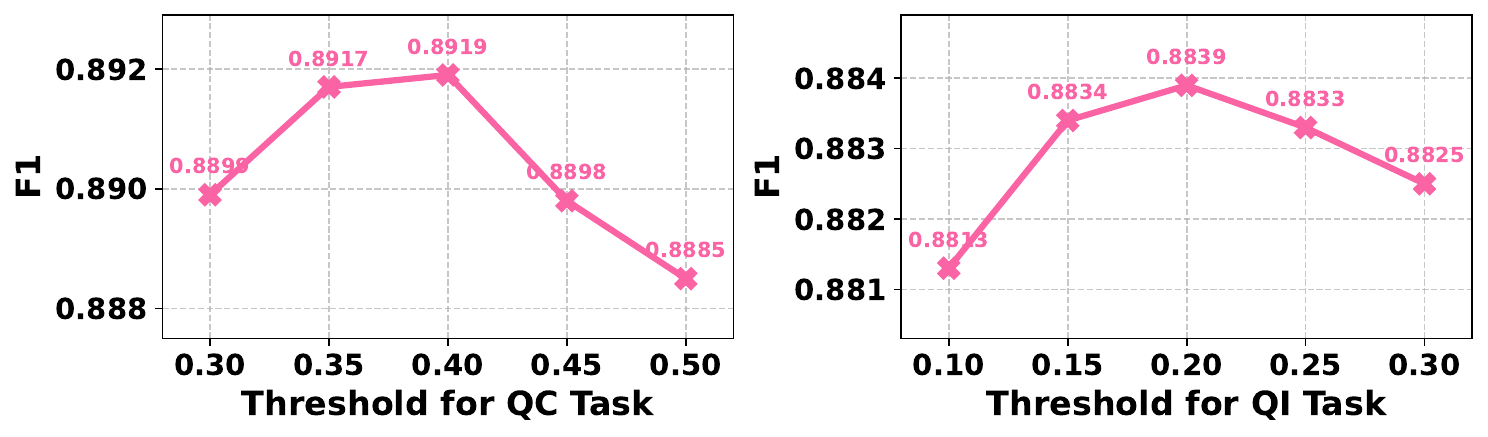}
\caption{
Performance comparison of QC and QI tasks across different prediction thresholds.
}
\label{fig:threshold}
\end{figure}
Our analysis reveals distinct optimal threshold ranges for the two tasks, validating the need for task-specific calibration. As shown in Figure \ref{fig:threshold}, the QI task achieves peak performance at thresholds between 0.15-0.2, while the QC task performs best at a significantly higher range of 0.35-0.4. These results validate our design choice of employing task-specific thresholds rather than a uniform cutoff. The divergence in optimal thresholds reflects fundamental differences in data semantics and label distributions between QC and QI tasks. More importantly, this threshold tuning enables additional performance gains that are independent of model training, demonstrating that inference-stage optimization can serve as an effective complement to data-centric approaches.

\subsection{Final Test Results}

\begin{table}[h]
\centering
\caption{Final Online Competition Results (Top 3 Teams)}
\label{tab:final_results}
\begin{tabular}{c|l|ccc}
\toprule
\textbf{Rank} & \textbf{Team} & \textbf{Overall} & \textbf{QC} & \textbf{QI} \\
\midrule
1 & DcuRAGONS & \textbf{0.8931} & \textbf{0.8965} & \textbf{0.8897} \\
2 & Industry\_AI & 0.8889 & 0.8928 & 0.8851 \\
3 & \textbf{Ours} & 0.8865 & 0.8896 & 0.8833 \\
\bottomrule
\end{tabular}
\end{table}

Our approach achieved third place in the final evaluation (Table \ref{tab:final_results}), demonstrating significant improvements over baseline methods while highlighting the effectiveness of pure data-centric strategies. The results validate our framework's capability to address multilingual generalization challenges through systematic data engineering.

Although a performance gap persists, our result is notable for being achieved by a purely data-centric strategy. This offers two key implications: (1) it validates data engineering as a highly effective and often more accessible lever for performance gains, and (2) it positions our method as a promising component that could be integrated with other advanced techniques to potentially achieve state-of-the-art results.

\section{Conclusion and Future Work}
This paper presents a data-centric framework to overcome key challenges in multilingual e-commerce search for Query-Category and Query-Item relevance. By systematically addressing linguistic disparity, class imbalance, and annotation noise through three core strategies—translation-based augmentation, semantic negative sampling, and self-validation filtering—we demonstrate that significant performance gains are achievable through data engineering alone, without modifying the underlying model architecture. Extensive validation on the CIKM AnalytiCup 2025 competition dataset shows our approach consistently outperforms strong baselines, achieving competitive results. The task-dependent effectiveness of different strategies provides valuable insights for practitioners, emphasizing that diagnostic analysis of data characteristics should precede enhancement implementation. This work underscores data-centric engineering as a potent and model-agnostic pathway for advancing real-world multilingual search systems.

Looking forward, several promising research directions emerge. First, enhancing robustness to query variations through targeted data augmentation—such as generating realistic misspellings or leveraging phonetic algorithms—could improve resilience against spelling errors and slang. Second, formalizing iterative data refinement through human-in-the-loop pipelines could create a virtuous cycle of continuous dataset improvement. Third, integrating external knowledge sources like product knowledge graphs could provide powerful semantic signals for resolving ambiguities in complex queries. Additionally, exploring advanced reinforcement learning techniques tailored for this domain, and experimenting with larger, more specialized foundation models present exciting opportunities for further advancing multilingual product search systems.

\bibliographystyle{ACM-Reference-Format}
\bibliography{main}







\end{document}